\documentclass[twocolumn,showpacs,prb]{revtex4}

\usepackage{psfrag}
\usepackage{enumerate}
\usepackage{graphicx}
\usepackage{amssymb}
\usepackage[tbtags]{amsmath}

\begin{document}

\title{Absence of a structural glass phase in a monoatomic model liquid
predicted to undergo an ideal glass transition}

\author{Charlotte Gils}
\author{Helmut G.~Katzgraber}
\author{Matthias Troyer}

\affiliation
{Theoretische Physik, ETH Z\"urich,
CH-8093 Z\"urich, Switzerland}

\date{\today}

\begin{abstract}

We study numerically a two-dimensional monodisperse model of
interacting classical particles predicted to exhibit a static
liquid-glass transition. Using
a dynamical Monte Carlo method we show that the model does not freeze
into a glassy phase at low temperatures. Instead, depending on the
choice of the hard-core radius for the particles, the system either
collapses trivially or a polycrystalline hexagonal structure emerges.

\end{abstract}

\maketitle

\section{Introduction}
\label{sec:intro}

There has been considerable progress in the physics of structural
glasses in the last few years.\cite{angell:95,richert:98} While
experimentally as well as numerically new insights have been gained,
theoretical studies of {\em simple} glass-forming models lag behind.
A structural glass is a solid state characterized by `random'
particle positions such that it is not possible to identify symmetries,
i.e., no long-range order exists. The nature of the transition from a
liquid to such a disordered phase is still an open question (see for
example Refs.~\onlinecite{kob:02} and \onlinecite{schilling:05}).
A popular microscopic approach is that of mode-coupling theory which
predicts a purely dynamical transition, i.e., a change in  the
dynamics from ergodic to non-ergodic behavior (see for example
Ref.~\onlinecite{goetze:92}).  In contrast, a static glass transition
arises from an equilibrium classical statistical mechanics approach
using a replica formulation,\cite{mezard:99a,mezard:99} based on
the assumption of an entropy crisis scenario.\cite{gibbs:57} It is
important to note that this approach assumes that the system is in a
uniform phase, thereby ignoring a crystal state from the very outset.
A static glass transition may therefore only arise in systems where
a crystal state does not dominate the low-temperature phase, e.g.,
by being separated from the glass phase by a large energy barrier.

\begin{figure}[!tbp]
\includegraphics[width=7.5cm]{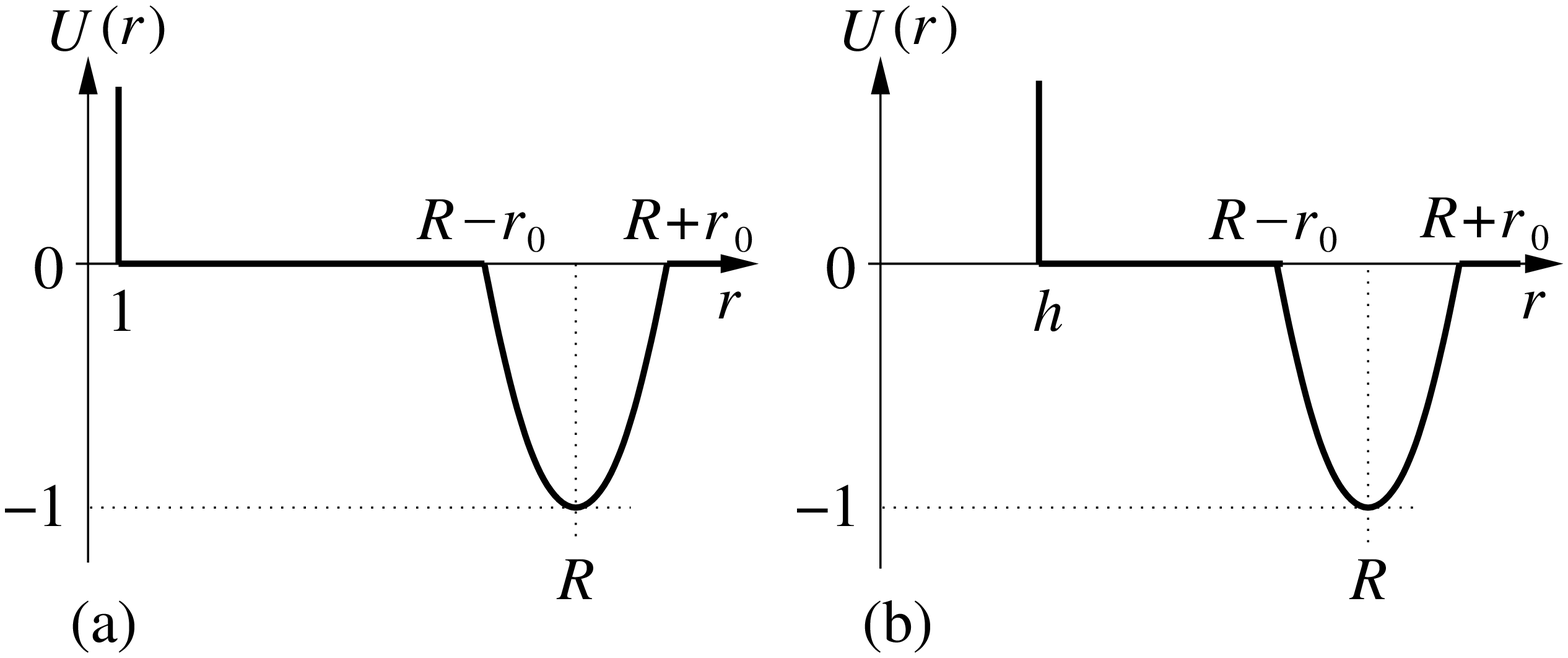}
\vspace*{-0.2cm}
\caption{
Schematic plot of the interaction potential $U(r)$
[Eq.~(\ref{eq:potential})] between two particles with relative distance $r
=|{\bf r}_i - {\bf r}_j|$.  For $r\in [R-r_0,R+r_0]$ the particles feel
an attraction, otherwise the potential is zero. Note that the model
is defined in the limit $1\ll r_0 \ll R$. (a) Interaction potential as
introduced in Ref.~\onlinecite{dotsenko:05} with a hardcore radius $h
= 1$. (b) Modified interaction potential in order to avoid a trivial
particle collapse with a hardcore radius $R/4<h<R/2$.
}
\label{fig:pot}
\end{figure}

The replica technique  has been applied to a variety of systems,
such as identical hard spheres,\cite{parisi:05}  a Lennard-Jones
binary mixture\cite{coluzzi:00} and to a monoatomic model with
attractive two-body interactions.\cite{dotsenko:05} The last study
has the advantage over conventional glass-forming models in that {\em
exact} replica calculations can be performed thus potentially being
able to deliver new insights into this complex field.  The authors
of Ref.~\onlinecite{dotsenko:05} consider a system of $N$ classical
particles in $D$ dimensions which interact by a two-body potential
which is nonzero for a given distance $R$ between the particles and
zero otherwise (see Fig.~\ref{fig:pot}).

Analytical\cite{moore:05a,tarzia:07} and numerical
studies\cite{rintoul:96} report the absence of a static glass
phase in monodisperse hard spheres in two and three dimensions,
respectively.\cite{comment:polydisperse}  Therefore it would be of
interest to test if the model proposed in Ref.~\onlinecite{dotsenko:05}
has a static glass transition even though the model is intrinsically
monodisperse [see Eq.~(\ref{eq:potential}) below]. In this work we
study the model proposed by Dotsenko and Blatter\cite{dotsenko:05}
numerically in two space dimensions ($D = 2$) and find that for
various combinations of the model parameters the system crystallizes
immediately.  We thus conclude that some amount of disorder must be
introduced in order to stabilize the glass phase.

\section{Model}
\label{sec:model}

We study the monoatomic glass former proposed in
Ref.~\onlinecite{dotsenko:05}.  The model describes $N$ particles in
$D$ space dimensions with a two-body interaction potential
\begin{equation}
U = \sum_{i<j} U( {\bf r}_i,{\bf r}_j) \, ,
\label{eq:energy}
\end{equation}
where 
\begin{equation}
U({\bf r}_i,{\bf r}_j) = 
	\frac{1}{r_0^2} \left(|{\bf r}_i - {\bf r}_j|-R\right)^2 - 1
\label{eq:potential}
\end{equation} 
for $R_- <|{\bf r}_i - {\bf r}_j| < R_+$.  Here $R_{\pm} = R \pm r_0$,
where $R$ is the `interaction radius' of the potential and $r_0$ the
width of the parabolic potential well [see Fig.~\ref{fig:pot} for
details].  The particle size is set to unity, which is equivalent to
a hardcore radius of $h = 1$. In these units the results obtained in a
mean-field approximation are valid in the parameter regimes $1 \ll r_0
\ll R$ and for an average particle distance  which equals approximately
those of the crystal and the liquid phase, i.e. $\rho^{-1/D} \approx
R$ ($\rho$ denotes the density).  The liquid and glass phases are
predicted to be separated\cite{dotsenko:05} by a critical temperature
\begin{equation}
T_{\rm c} \sim 1/\ln(DR/r_0) \, .
\label{eq:tc}
\end{equation}
The disordered nature of the phase for temperatures below $T_{\rm c}$
is identified by means of a replica correlator which quantifies the
correlations between particles in two or more replicas.

\begin{figure}[!tbp]
\includegraphics[width=8.5cm]{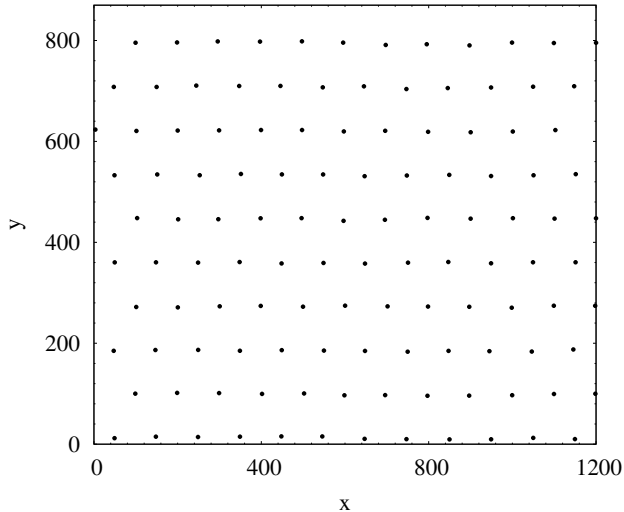}
\caption{
Configuration of particles after $2\cdot 10^8$ Monte Carlo steps
for $r_0=10$, $R=100$, $h=30$ (i.e., $T_c \approx 0.33$), $T=0.1$,
$N=120$ and an average particle spacing $\rho^{-1/2} \approx 88.3$.
}
\label{fig1}
\end{figure}

\section{Numerical Method and Results}
\label{sec:results}

We use a Metropolis Monte Carlo (MC) method (see for example
Ref.~\onlinecite{krauth:98} for details).  Without loss of generality
we study particles in two space dimensions with periodic boundary
conditions in a box of size $L_x \times L_y$.  A MC step consists of
choosing a new configuration of particle positions and accepting this
move with probability $p = \min[1,\exp(-\Delta U/T)]$, where $\Delta U$
is the difference in the potential  energy  between the old and new
configuration of particles.  Two types of local moves are applied: a
random particle and a random direction are chosen, and the particle is
moved into this direction by a distance uniformly drawn from either the
interval $[0,r_0]$, or the interval $[R_-,R_+]$. The resulting local
dynamics serves as a model for the realistic dynamics of the system
in renormalized physical time.  In the absence of ergodicity breaking,
the method also allows one  to study the thermodynamic phase of the system:
after a sufficient number of MC steps configurations are distributed
according to the equilibrium distribution $\exp(-U/T)$.  Clearly, if
no disordered phase is found after relatively few MC steps, the use
of optimized nonlocal MC updates which allow the system to assume
low-energy configurations in a much faster manner will certainly
not lead to a disordered phase either.  The simulation is started
with a random configuration of particles. The temperature $T$ for
the simulation is chosen such that $T < T_{\rm c}$, where $T_{\rm
c}$ is given by Eq.~(\ref{eq:tc}).  Since the system starts from a
quenched configuration, the cooling rate is thus infinitely high,
as required in Ref.~\onlinecite{dotsenko:05}.

\begin{figure}[!tbp]
\includegraphics[width=8.5cm]{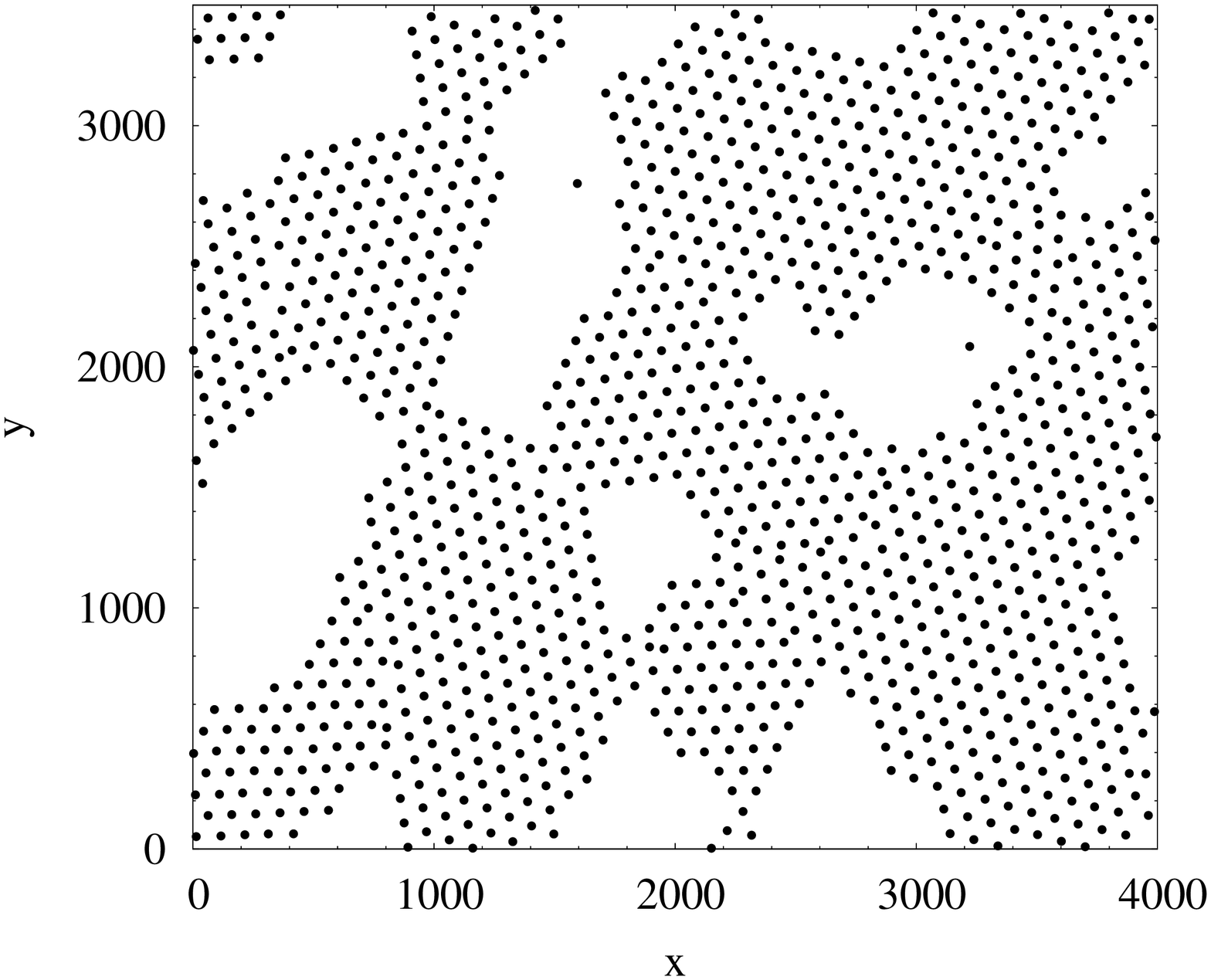}
\caption{
Configuration snapshot after $3\cdot 10^9$ Monte Carlo steps for
$r_0=10$, $R=100$, $h=30$, (i.e., $T_c \approx 0.33$), $T=0.1$,
$N=1200$ and an average particle distance $\rho^{-1/2} \approx 108$
(i.e., a smaller density than in Fig.~\ref{fig1}).  A polycrystalline
structure with randomly-oriented patches of hexagonal crystal is found.
}
\label{fig2}
\end{figure}

For the glass phase to be stable Dotsenko and Blatter\cite{dotsenko:05}
give clear restrictions on the parameters of the model
glass former, i.e., $1\ll r_0 \ll R$. Simulating the system with
the standard hard-core radius $h = 1$ and for temperatures $T <
T_{\rm c}$ yields a trivial collapse where all particles approach
each other up to a minimum distance of $2h$. We have tested this
scenario for various choices of all system parameters.  The same
trivial collapse is  found for any hard-core radius which satisfies
$h < R/4$.  To avoid this behavior, we increase the size of the
hardcore radius to values $R/2 >h >R/4$. Note that the size of the
hard-code radius $h$ is irrelevant in the analytic calculations of
Ref.~\onlinecite{dotsenko:05}.\cite{dotsenko:h}

\begin{figure*}[!tbp]

\includegraphics[width=8.5cm]{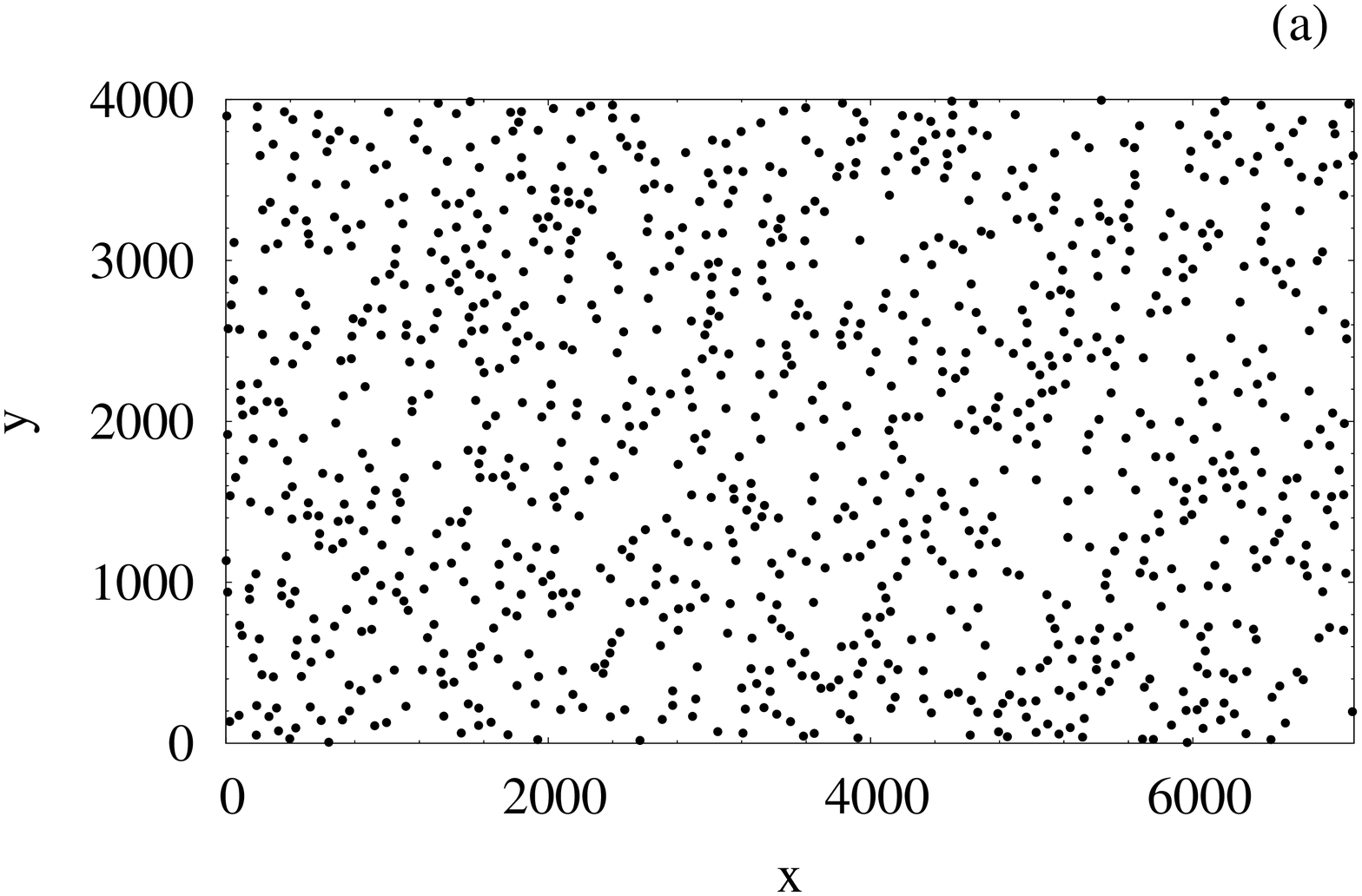}
\includegraphics[width=8.5cm]{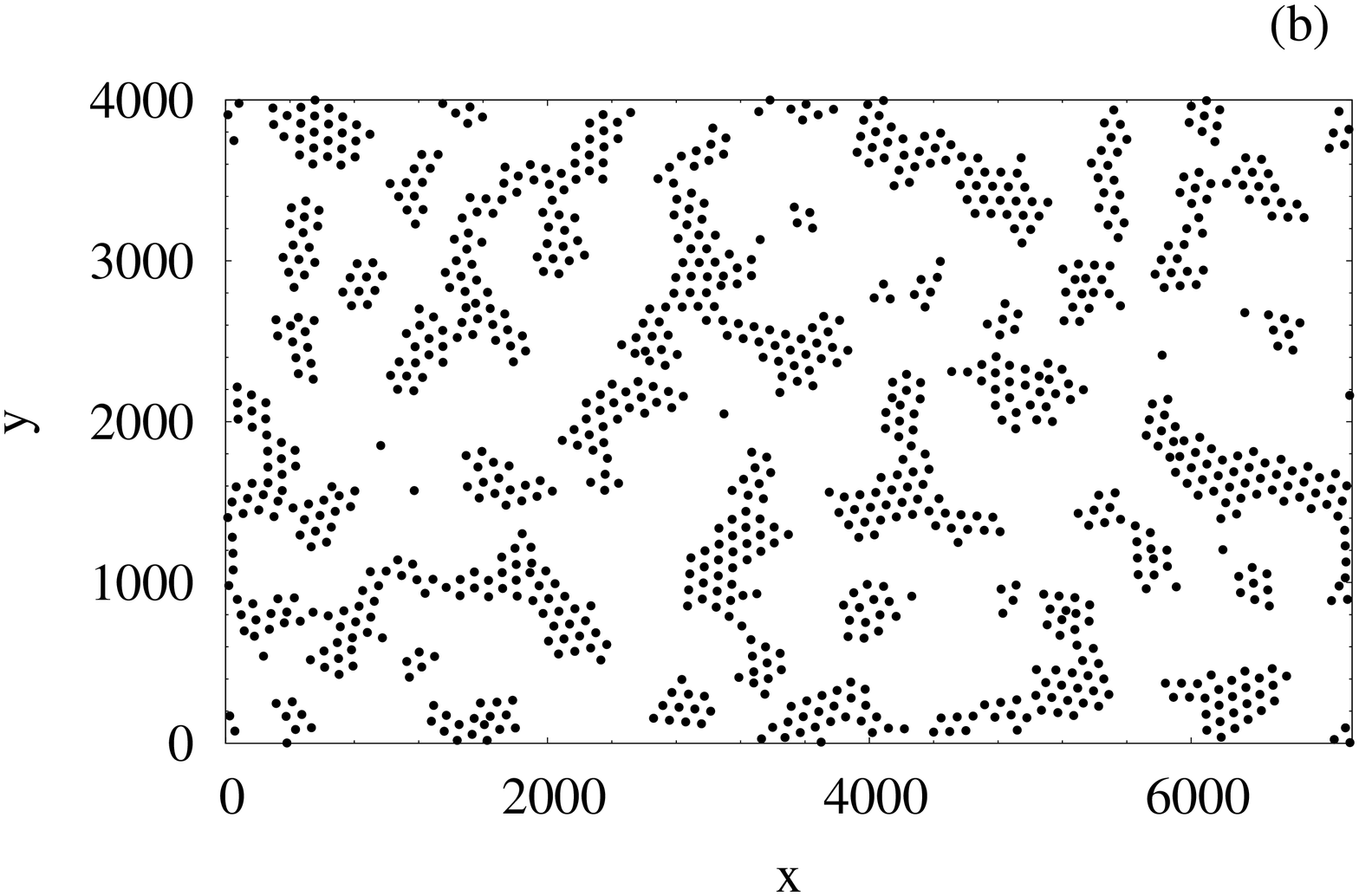}

\includegraphics[width=8.5cm]{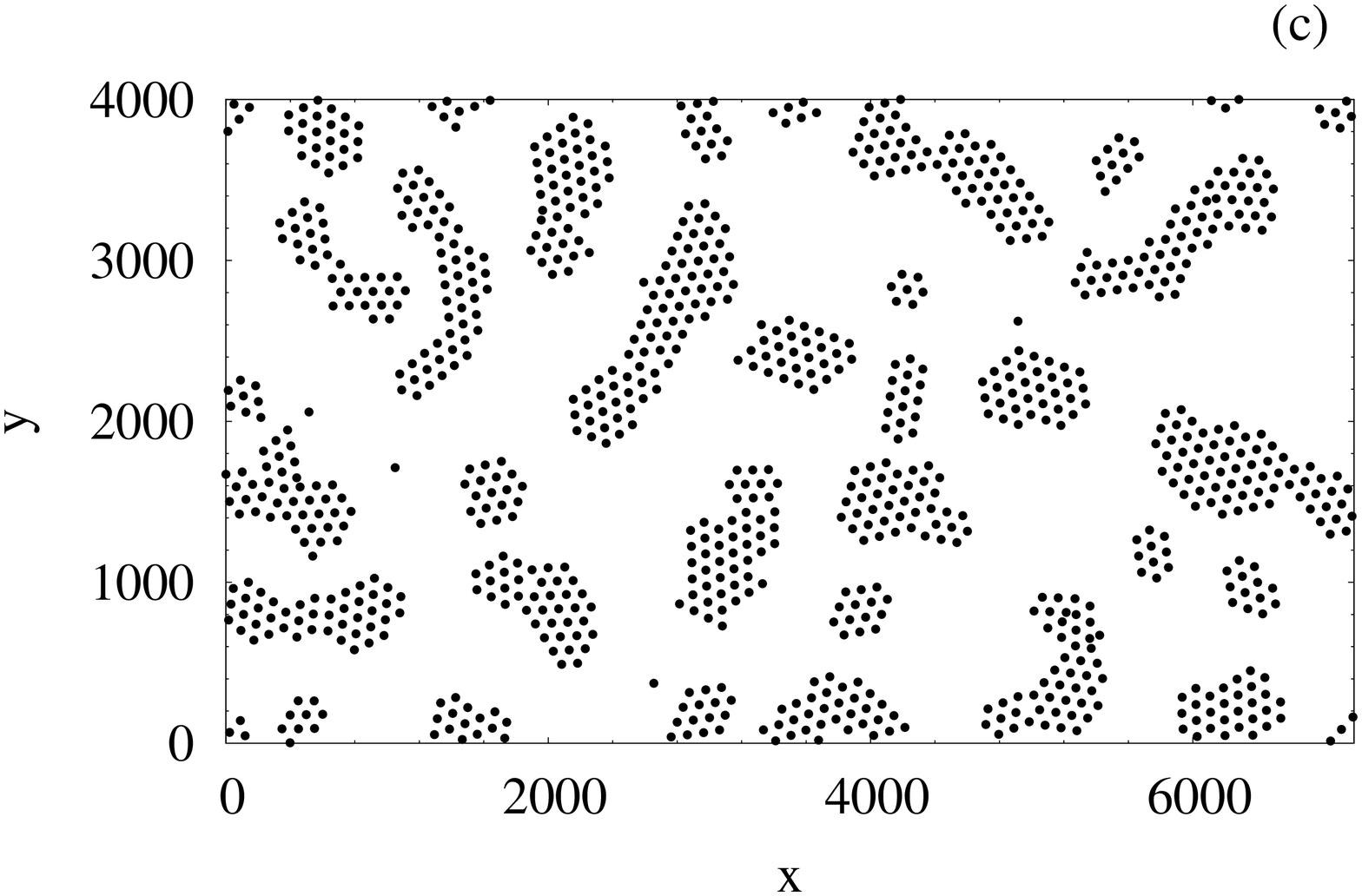}
\includegraphics[width=8.5cm]{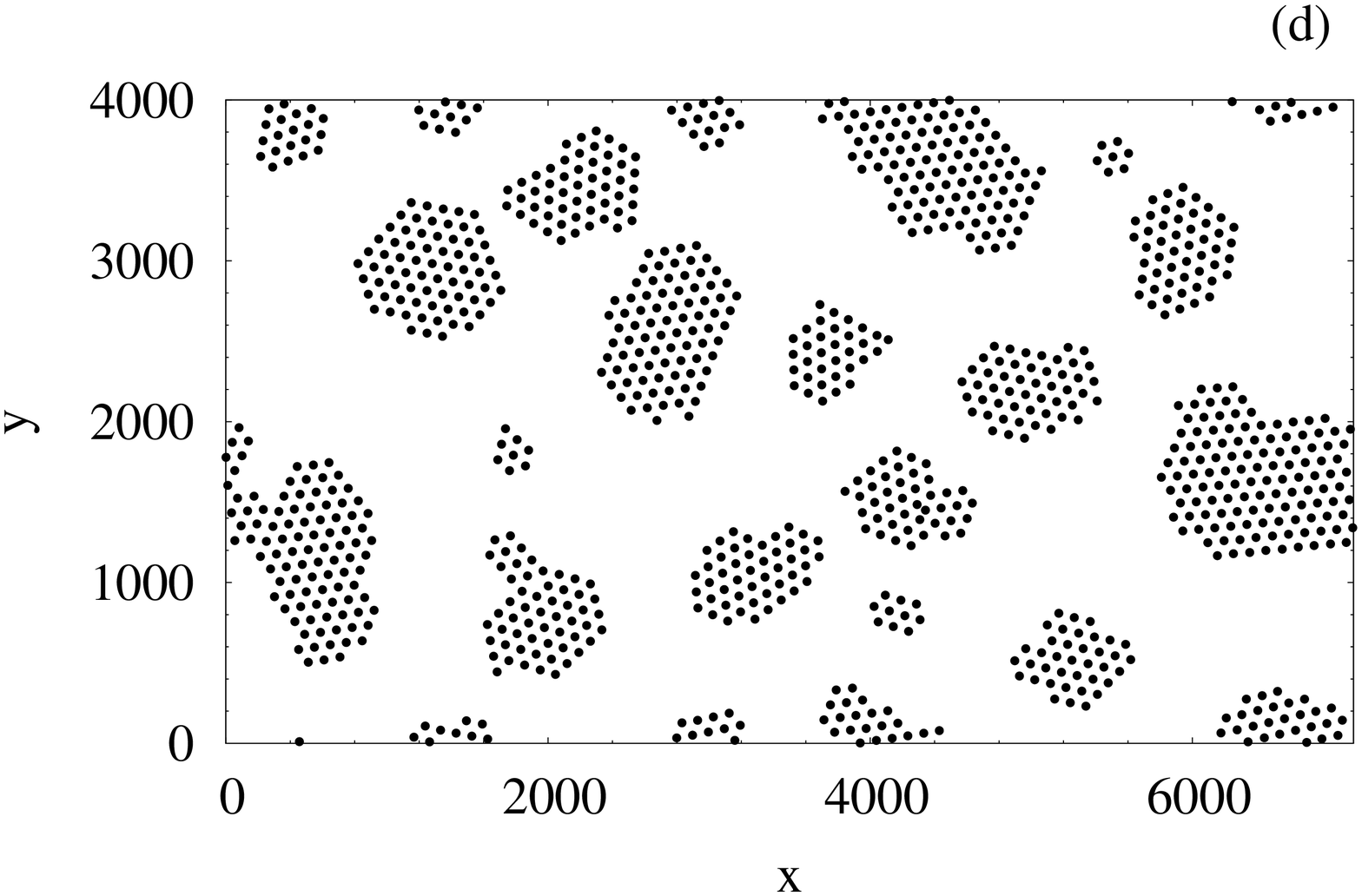}

\vspace*{-0.2cm}

\caption{ 
Particle configurations for $N=1000$ particles ($r_0 = 10$, $R =
100$, $h = 30$, $T=0.05 \ll T_c$) in a system with
average particle distance $\rho^{-1/2} \approx 167$.  (a) Initial
random configuration, (b) configuration snapshot after $2 \cdot 10^7$
MC steps, (c) configuration snapshot after $2 \cdot 10^8$ MC steps,
and (d) after $5 \cdot 10^9$ MC steps.  As in Fig.~\ref{fig2},
a polycrystalline structure emerges.
}
\label{fig3}
\end{figure*}

For the modified interaction potential [Fig.~\ref{fig:pot}(b)] a
hexagonal crystal structure emerges: If $L_x$, $L_y$ and $N$ are chosen
appropriately, a perfect crystal develops as shown in Fig.~\ref{fig1}.
If the density is lower than the density of the hexagonal crystal,
the particles organize in a polycrystalline structure: small hexagonal
crystal patches with random orientations form, as can be seen in
Fig.~\ref{fig2} and Fig.~\ref{fig3}(d). In Figs.~\ref{fig3}(a)
-- \ref{fig3}(d) we show how the crystalline structure forms as
a function of time measured in Monte Carlo steps, starting from a
low-density random configuration [panel (a)].  We have also measured
the replica correlator [Eq.~(62) in Ref.~\onlinecite{dotsenko:05}]
whose finite value would be indicative of a glassy state. However,
in our simulations, it takes a large value close to the linear system
size squared, which means that particles in two replicas are completely
uncorrelated (not shown). For large systems and various densities,
ranging from those of the crystal state [Fig.~\ref{fig1}] to much more
dilute systems [Figs.~\ref{fig2} and ~\ref{fig3}], our results are in
stark contrast to the predictions of Ref.~\onlinecite{dotsenko:05}:
While a glassy ``network-like'' structure with an average number of
nearest-neighbors which is significantly smaller than in a crystalline
state is predicted, our results are locally crystal-like and show
a high degree of order.\cite{comment:sf,comment:wk} We have varied
the system parameters, in particular increasing the ratio of $R$ and
$r_0$, but the results have been qualitatively similar to the ones
shown in the Figures. Therefore, if a glass phase exists in the model,
it cannot be observed since crystallization takes place very quickly.

\section{Conclusions}
\label{sec:conclusions}

In conclusion, our results from Monte Carlo simulations are in
agreement with previous predictions\cite{moore:05a,tarzia:07} that
monoatomic systems in two space dimensions with two-body interactions,
such as the recent proposal\cite{dotsenko:05} studied here in
detail, cannot form a glassy structure at low temperatures. Instead,
the system forms a polycrystalline structure and does not show
an ideal liquid-glass transition.  This is unfortunate since
the proposed monoatomic model is analytically solvable and thus
holds the promise to deliver new insights to the physics of glass
formers. It might be of interest to verify if the model is still
analytically solvable if, for example, many-body interactions between
the particles\cite{dileonardo:84} are added to the Hamiltonian or
Gaussian randomness is introduced into the distance $R$ to suppress
the crystal state and favor a glassy phase.  Assuming the crystal
structure is better suppressed in higher space dimensions, it might
be conceivable to find a glass structure for $D > 2$ and thus possibly
determine a lower critical dimension of the model.

While the simulations are performed for finite systems, we believe
the number of particles studied shall not influence the conclusions
found. In particular, we have scanned the vast parameter space in
detail and have performed simulations for many different particle
numbers and densities always yielding similar results.

\begin{acknowledgments}

We would like to thank G.~Blatter, V.~S.~Dotsenko and W.~Krauth for
helpful discussions. The simulations were performed on the hreidar
cluster at ETH Z\"urich. H.G.K.~acknowledges support from the Swiss
National Science Foundation under grant No.~PP002-114713.

\end{acknowledgments}

\bibliography{comments,refs}

\begin{thebibliography}{22}
\bibitem[{com({\natexlab{a}})}]{comment:polydisperse}
\bibinfo{note}{Note that for polydisperse systems the existence and nature of
  the glass phase is still controversial (see Refs.~\onlinecite{santen:00} and
  \onlinecite{brumer:04}).}
\bibitem[{com({\natexlab{b}})}]{comment:sf}
\bibinfo{note}{We have also measured the structure factor of the system which
  suggests an underlying crystalline structure (peaks) and does not show the
  typical ring-like feature found for a disordered system such as a glass or
  liquid.}
\bibitem[{com({\natexlab{c}})}]{comment:wk}
\bibinfo{note}{The probability of creating a crystal germ to overcome the
  surface energy cost grows linearly with system size. This means that for a
  large enough system a germ will form which eventually will spread over a
  macroscopic volume of the system (depending on the particle density it can be
  the whole system volume). Quoting Werner Krauth: ``Schwupp-di-wupp the whole
  box is a crystal''}.
\end{thebibliography}

\end{document}